\documentclass[prd,twocolumn,amsmath,amssymb,floatfix,nofootinbib,superscriptaddress]{revtex4}

\usepackage{graphicx}
\usepackage{bm}
\usepackage{amsmath,amssymb}
\usepackage{epsfig}
\usepackage{color}
\usepackage{natbib}
\usepackage{enumerate}
\usepackage[pdftex]{hyperref}
\usepackage{ifthen}
\usepackage{subfigure}
\usepackage{hhline}
\usepackage{multirow}
\pdfmapfile{+txfonts.map}

\newcommand{\PICO}{\textsc{PICO}}
\newcommand{\GetDist}{\textsc{GetDist}}

\newcommand{\planck}{\textit{Planck}}
\newcommand{\ndrag}{f_{\rm drag}}
\newcommand{\nslow}{N_{\rm slow}}
\newcommand{\nfast}{N_{\rm fast}}

\newcommand{\oversample}{f_{\rm fast}}

\providecommand{\CAMB}{\textsc{camb}}
\providecommand{\COSMOMC}{\textsc{CosmoMC}}

\providecommand{\LCDM}{{$\rm{\Lambda CDM}$}}

\newcommand{\begm}{\begin{pmatrix}}
\newcommand{\enm}{\end{pmatrix}}

\newcommand\ba{\begin{eqnarray}}
\newcommand\ea{\end{eqnarray}}
\newcommand\bea{\begin{eqnarray}}
\newcommand\eea{\end{eqnarray}}

\newcommand\be{\begin{equation}}
\newcommand\ee{\end{equation}}

\newcommand{\la}{\langle}
\newcommand{\ra}{\rangle}

\newcommand{\mC}{\bm{C}}
\newcommand{\mD}{\bm{D}}

\newcommand{\mL}{\bm{L}}
\newcommand{\mM}{\bm{M}}

\newcommand{\mU}{\bm{U}}

\newcommand{\boldvec}[1]{{{\mathbf{#1}}}}

\newcommand{\vF}{\boldvec{F}}

\newcommand{\vS}{\boldvec{S}}

\newcommand{\vx}{\boldvec{x}}

\newcommand{\clo}{\mathcal{O}}

\def\eprinttmp@#1arXiv:#2 [#3]#4@{
\ifthenelse{\equal{#3}{x}}{\href{http://arxiv.org/abs/#1}{#1}}{\href{http://arxiv.org/abs/#2}{arXiv:#2} [#3]}}

\providecommand{\eprint}[1]{\eprinttmp@#1arXiv: [x]@}
\newcommand{\adsurl}[1]{\href{#1}{ADS}}
\providecommand{\bibinfo}[2]{\ifthenelse{\equal{#1}{isbn}}{
\href{http://cosmologist.info/ISBN/#2}{#2}}{#2}}

\begin{document}

\title{Efficient sampling of fast and slow cosmological parameters}

\author{Antony Lewis}
\homepage{http://cosmologist.info}
\address{Department of Physics \& Astronomy, University of Sussex, Brighton BN1 9QH, UK}


\begin{abstract}
Physical parameters are often constrained from the data likelihoods using sampling methods. Changing some parameters can be much more computationally expensive (`slow') than changing other parameters (`fast parameters').
I describe a method for decorrelating fast and slow parameters so that parameter sampling in the full space becomes almost as efficient as sampling in the slow subspace when the covariance is well known and the distributions are simple. This gives a large reduction in computational cost when there are many fast parameters. The method can also be combined with a fast `dragging' method proposed by Neal~\cite{Neal04} that can be more robust and efficient when parameters cannot be fully decorrelated a priori or have more complicated dependencies. I illustrate these methods for the case of cosmological parameter estimation using data likelihoods from the \planck\ satellite observations with dozens of fast nuisance parameters, and demonstrate a speedup by a factor of five or more. In more complicated cases, especially where the fast subspace is very fast but complex or highly correlated, the fast-slow sampling methods can in principle give arbitrarily large performance gains. The new samplers are implemented in the latest version of the publicly available \COSMOMC\ code.
\end{abstract}
\maketitle

\vskip .2in
\section{Introduction}

Bayesian methods are often used to compare physical models to data, where parameters in different models are most easily constrained by sampling from the posterior distribution. Sampling methods scale well with the dimension of the parameter space, in the best case scaling linearly, compared to brute force integration which scales exponentially badly. However as the number of parameters in the models grow, the computational cost can still be challenging. This can happen either because the complexity of the physical model increases, or because, as the precision of the constraints increases, so does the complexity of the data analysis nuisance parameters.
This paper describes a way to improve on the scaling with dimension in the case where the the additional parameters are `fast' parameters. This reduces computational cost and also makes feasible the analysis of more realistic models with more parameters than might otherwise be easily tractable. The methods are general, but for concreteness I will focus on examples in cosmological parameter estimation that were the original motivation.

The separation of fast and slow cosmological parameters was introduced in Ref.~\cite{Lewis:2002ah}. The idea is that calculating a new likelihood when you change different parameters can have very different numerical cost depending on which parameters are changed. For example in cosmology changing a matter density parameter requires a full recalculation of the evolution of the background model and the perturbations, and then
all the data likelihoods that depend on them: this is slow. However changing a primordial power spectrum parameter does not change the linear transfer functions, so only the integrals to calculate the theoretical power spectra from the transfer functions need to be done. If the data likelihoods are fast, changing primordial power spectrum parameters is therefore much faster than changing density parameters.

Furthermore there are often many nuisance parameters. For example the data likelihood often depends on a variety of parameters governing the calibration, noise levels and other characteristics of the experiment. In cosmology, the likelihood from the \planck\ satellite observations has parameters governing foreground amplitudes and correlations, uncertainties in the instrument beam response, and calibration uncertainties. When these nuisance parameters are changed, keeping other physical parameters fixed,  only the dependent likelihood function has to be recomputed, not the theoretical predictions nor other independent likelihoods for other data being used. There can easily be a speed hierarchy of a hundred or more between changing the nuisance parameters and changing the main parameters governing the physical model. In some cases the nuisance parameters can be quickly numerically or analytically marginalized; however, in other cases they cannot, or the nuisance parameter posteriors are themselves of interest, so an efficient sampling method is required.

When there are many more fast parameters than slow parameters, there are potentially large efficiency gains to be had from exploiting the different computational speeds. If all parameters are sampled efficiently, and changing fast parameters is much faster than changing slow parameters, there is at least a linear saving (by a factor of order $1/f_{\rm slow}$, where $f_{\rm slow}$ is the fraction of parameters that are slow).
  However if exploration of the fast parameter space is difficult, or there are unmodelled degeneracies between fast and slow parameters, the saving could be arbitrarily larger than this.

I will focus in this paper on unimodal distributions that are nonpathological for good physical reasons, where simple standard Metropolis-Hastings~\cite{Metropolis53,Hastings70} sampling methods work well.
The outline of this paper is as follows: in Sec.~\ref{sec:cholesky} I describe a general method for constructing a set of
 decorrelated fast and slow parameters to allow efficient movement within the full parameter space. This can easily be used with standard Metropolis sampling, and also optionally with oversampling in the fast-parameter subspace for significant performance gains in some cases.
 In Sec.~\ref{sec:dragging} I describe an implementation of a fast-parameter `dragging scheme' introduced by Ref.~\cite{Neal04}. This can be used in combination with the parameter decorrelation and adaptive methods to provide efficiency gains in some non-Gaussian problems, or new problems with strong fast-slow correlations where the correlation structure is not well known a priori. In Sec.~\ref{sec:performance} I discuss the merits and scaling of the different methods in some simple limiting cases and a useful convergence measure, and then go on to discuss indicative performance in the specific realistic case of parameter inference from \planck\ satellite observations (for which these methods were originally implemented). I finish with conclusions, and discuss in the Appendix a few more general details of the proposal distribution implemented in the public \COSMOMC\ code\footnote{\url{http://cosmologist.info/cosmomc/}}.

\section{Fast-slow decorrelation}
\label{sec:cholesky}

The performance of Metropolis-Hastings Markov-chain Monte-Carlo (MCMC) sampling methods depend strongly on the shape of the distribution and also on the choice of proposal distribution.
If there are correlated parameters a good choice of proposal distribution will propose longer steps along the degeneracy directions. For simple distributions this is first done by intelligently choosing the base parameters~\cite{Kosowsky:2002zt} to remove tight non-linear degeneracies.
Then using an estimate of the covariance of the base parameters (e.g. from an initial run, or the first steps of an adaptive method)
by orthogonalizing the base parameters, so that proposals are made to linear combinations of the original base parameters that are nearly independent. This allows rapid movement through the distribution, and hence fewer steps between independent samples and faster convergence to the desired distribution.

The simplest method to exploit fast parameters is to make separate proposals sequentially is the fast and slow parameter subspaces~ \cite{Lewis:2002ah}. However a potential problem is that fast and slow parameters can be arbitrarily correlated. We want to redefine parameters to be as uncorrelated as possible, but in such a way that that fast and slow parameters are not all mixed together, so that changing a subset of the new  parameters remains fast.

\begin{figure}[t]
\begin{center}
\includegraphics[width=\hsize]{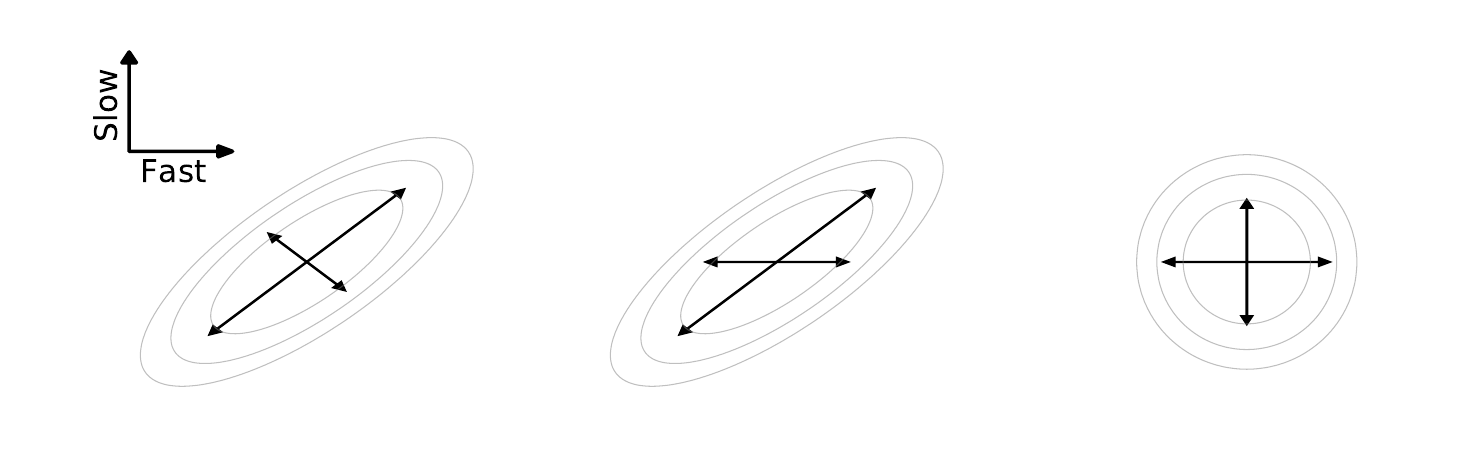}
\caption{Possible proposal directions (arrows) for sampling a correlated 2D distribution:
\emph{Left:} a choice of orthogonal eigenvector directions explores efficiently but requires changing both fast and slow parameters in both proposal directions; \emph{Centre:} a choice that allows fast moves in the $x$-direction but is non-orthogonal;
\emph{Right:} By performing a linear shearing (parameter redefinition of the slow direction) the proposal distribution can be orthogonal \emph{and} changes in the fast $x$ direction can remain fast.\label{fastslow}}
\end{center}
\end{figure}

Fortunately it is possible to simultaneously decorrelate all the parameters, and maintain the same number of fast directions in parameter space. The idea is illustrated in Fig.~\ref{fastslow}.
To do this we first order the parameters by speed, so that if $i<j$ then $x_i$ is slower than (or the same speed as) $x_j$.
We can now make a linear parameter redefinition, so that the new slow parameters depend on the original fast and slow parameters, but the new fast parameters do not depend on slow parameters. To do this Cholesky decompose the covariance as
\be
\mC = \la \vx \vx^T\ra = \mL \mL^T,
\ee
where $\mL$ is a lower triangular matrix. Then the new decorrelated parameters are taken to be $\vx' = \mL^{-1} \vx$.
If a proposed move in the new orthonormal $\vx'$ space is $\vx' \rightarrow \vx'+\Delta \vx'$, then in the original space
\be
\vx \rightarrow \vx + \mL \Delta\vx'.
\ee
The lower triangular nature of $\mL$ ensures that if $\Delta\vx'$ only has non-zero components with $i\ge j$, then parameters in $\vx$ are also only modified where $x_i$ has $i\ge j$. Therefore, the decorrelated parameters $\vx'$ are also ordered by speed, in that changing the $i$th decorrelated parameter only requires calculating likelihood changes for parameters which are as fast or faster than\footnote{Of course this typically will be slower than just changing $x_i$, which is the price paid for decorrelation. If certain blocks of parameters are known to be only slightly correlated, their independence could be imposed so that the changing one does not require changing parameters in the other block.} $x_i$.

To see the structure in a simple example, consider
a parameter vector consisting of two slow parameters $S_i$ and several fast parameters $F_i$. The parameter vector $\vx^T = (S_1 S_2 F_1 F_2 F_3 \dots)$ is then related to the orthonormalized parameters $\vx'$ by
\setlength{\arrayrulewidth}{.5pt}
\be
\begm
S_1 \\ S_2 \\  F_1 \\ F_2 \\ F_3 \\ \vdots
\enm
=\left[
\begin{array}{cc|cccc}
* & 0 & 0 & 0 & 0 &\dots \\
* & * & 0 & 0 & 0 &\dots \\ \cline{3-6}
* & * & * & 0 & 0 &\dots \\
* & * & * & * & 0 & \dots \\
* & * & * & * & * & \dots \\
\vdots &\vdots &\vdots &\vdots &\vdots &
\end{array}\right]
\begm
S_1' \\ S_2' \\  F_1' \\ F_2' \\ F_3' \\ \vdots
\enm,
\ee
where stars denote the non-zero elements of the lower triangular matrix $L$ from the Cholesky decomposition of the covariance.
This is of the form
\be
\begm
\multirow{3}{*}{$\vS$} \\  \\ \cline{1-1} \multirow{3}{*}{$\vF$} \\ \\ \\ \\
\enm
=\left[
\begin{array}{cc|cccc}
 &  & 0 & 0 & 0 &\dots\\
 &  & 0 & 0 & 0 & \dots \\ \cline{3-6}
 &  &   &  &   &  \\
 \multicolumn{2}{c|}{\mM_S}  &  &  &  &\\
 &  &  &  \multicolumn{2}{c}{\mM_F}   &  \\
 &  &  &  &  &
\end{array}\right]
\begm
\multirow{3}{*}{$\vS'$} \\  \\ \cline{1-1} \\ \multirow{3}{*}{$\vF'$}  \\ \\ \\
\enm,
\ee
where the $\mM_S$ and $\mM_F$ submatrices can be precomputed from an estimate of the covariance.
Under a proposed move in the slow orthonormalized subspace $\Delta \vx' = (\Delta\vS', 0)$ we then have
\be
\Delta \vx = \mM_S \Delta \vS',
\ee
so all of the original parameters are changed. However, under a move in the fast orthonormalized subspace $\Delta \vx' = (0, \Delta\vF')$, only fast parameters change:  $\Delta \vS=0$ and
\be
\Delta \vF = \mM_F \Delta\vF'.
\ee
This can be generalized to many blocks of fast and slow parameters of similar speed. Standard Metropolis-Hastings moves can now be made in the diagonalized fast-slow parameter space, and only moves along slow directions will be slow. This achieves fast movement in the slow parameter space, while maintaining efficiency gains from exploiting the fast-slow decomposition.

\subsection{Block randomization}

Proposing sequentially in the $\vx'$ parameters allows for efficient exploration of parameter space, with changes to fast parameters remaining fast. However, we would also like our sampler to be a bit robust; for example, if the assumed covariance is not quite right, or the shape of the posterior is non-Gaussian. The choice of proposal distribution function is discussed in Appendix~\ref{proposal}. In particular it is usually better to sample in random directions in $\vx'$ space, rather than along the axes. The approach used in \COSMOMC\ is to sample cyclically from a randomly rotated coordinate basis (with the random rotation periodically changed, e.g. once per complete cycle). This allows all directions to be eventually proposed, and also improves robustness to the covariance in one direction being badly misestimated.

With fast and slow parameters this is not a good idea because it would mix fast and slow parameters in each proposal. Instead, parameters can be blocked into groups of equal or similar speed, with the sampler making proposals along random directions in these equal-speed subspaces. This will be slightly less robust than using random directions in the full space, but perfectly preserves the fast-slow parameter separation. This is what is implemented in \COSMOMC\, along with random ordering of proposals between different blocks.

\subsection{Fast-parameter oversampling}

Given a set of parameter blocks of similar speed, we are free to choose the relative number of proposals made within each block. If all parameters are equivalent, the most efficient symmetric solution is to sample all parameter directions equally (i.e., each block is sampled proportional to the number of parameters in it). However changing the fast parameters is fast, so it may be advantageous to make more steps in the fast blocks than in the slow blocks. I define a parameter $\oversample$ so that $\oversample$ times more fast parameter proposals are made than under the equal-sampling scheme; the total number of proposals per cycle is then $\nslow + \oversample\nfast$.
In general there could be many different speed blocks, and each could have its own $\oversample$ factor, but for simplicity I restrict to just one. Using $\oversample>1$ will in general be beneficial in that it allows better exploration of the fast subspace for small additional computational cost. The only cases where it is not likely to be beneficial is when it is known that the fast subspace is not actually that fast, the fast subspace is much easier to explore than the orthogonalized slow parameter space, or the fast parameters are nearly independent of the slow parameters and the only goal is to have small sampling error on functions of the slow parameters.

Using $\oversample\gg 1$ will allow fast parameters to fully explore the conditional distribution for each point in the slow parameter space. This can be a large gain if this subspace is hard to explore, and in general helps to reduce to sample variance fluctuations. However, it will not greatly improve convergence in the full parameter space if there are correlations with the slow parameters, since the overall sampling efficiency is still limited by the efficiency with which the slow parameter space is explored. The method described in the next section describes an alternative method that uses many fast likelihood evaluations to improve the movement in the slow parameter space, which for distributions with general fast-slow dependencies and a large speed hierarchy can be more efficient.

\section{Dragging fast variables}
\label{sec:dragging}

In Ref.~\cite{Neal04} Neal devised a general scheme for sampling fast and slow parameters by `dragging' the fast parameters along each slow Metropolis proposal. The idea is that when varying slow parameters ideally we would like to sample the fully fast-marginalized probability, so that there is no random walk behaviour from exploring correlations with fast parameters or funny shapes in the full parameter space. For Gaussian distributions the decorrelation scheme described in Sec.~\ref{sec:cholesky} will of course achieve this anyway since the marginalized and conditional distributions are the same in a fully orthonormalized parameter space. However, in practice, the distributions are not Gaussian, and there may be only an approximate covariance to do the decorrelation, so any scheme that samples from the fast-marginalized distribution efficiently will be more robust. Neal's dragging method asymptotically achieves this by a sampling method.

\begin{figure}[t]
\begin{center}
\includegraphics[width=\hsize]{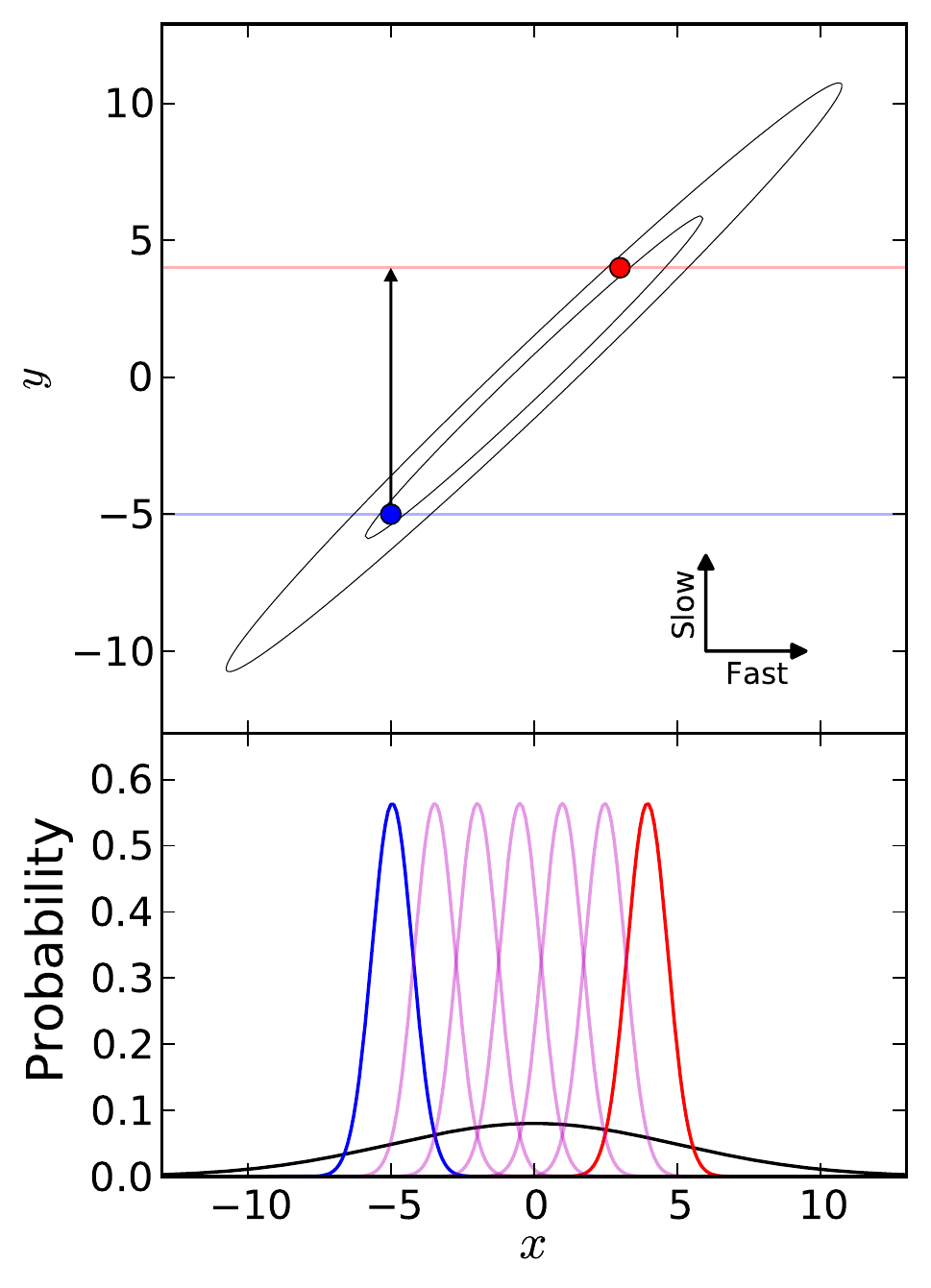}
\caption{Illustration of sampling from a correlated 2D distribution (contours, where the correlation is unknown a priori and hence not taken out by a variable decorrelation transformation). Here $x$ is a fast direction, $y$ is slow. Consider proposing a move $y\rightarrow y'$ as shown by the arrow in the top plot. Normally this would be immediately rejected with very high probability. \emph{Lower plot:} the dragging method takes samples in the fast $x$ direction from a series of interpolating distributions [magenta lines] that interpolate between $P(x|y)$ [blue] and $P(x|y')$ [red]. This allows the fast samples to gradually explore the degeneracy direction, for example ending up at the red point in the upper plot. If there are enough interpolating steps the total move is accepted with probability similar to sampling from the marginalized distribution $P(x)$ [solid black line in the lower plot], which in general is not possible directly.
\label{dragging}}
\end{center}
\end{figure}

The method works by making a proposal in the slow parameter space, and then running a chain in the fast parameter space
that is guided to explore any fast-slow degeneracy and move towards the region of the fast parameter space that has high likelihood at the proposed slow parameter point. By suitably sampling from distributions that interpolate between those at the two end points, the fast parameters can be `dragged' along any degeneracy direction, and the acceptance probability for the full slow move approaches that expected from sampling directly from the fast-marginalized distribution (which you can't actually calculate for large numbers of fast parameters), see Fig.~\ref{dragging}.

The method is a variant of the Metropolis-Hastings algorithm, where at each point in the chain a move is proposed and then accepted depending on a probability ratio. Sampling from a distribution $P(x,y)$ depending on $\nfast$ fast parameters (direction $x$) and $\nslow$ slow parameters (direction $y$) the basic steps for symmetric proposal distributions are as follows~\cite{Neal04} :
\begin{enumerate}
\item From the current slow parameters $y$ propose a new set of slow parameters $y'$
\item Construct a series of $n-1$ distributions $P_i(x)$ that interpolate between $P(x|y)$ and $P(x|y')$
with
\be
\ln P_i(x) \equiv \frac{ (n-i) \ln P(x|y) +i \ln P(x|y')}{n}
\ee
for $i=0,\dots, n$. With this definition $P_0(x)\equiv P(x|y)$ at the starting slow position and $P_n(x)\equiv P(x|y')$ at the proposed slow position.
\item Sample a chain of fast parameters $x$ by in turn sampling from $P_i(x)$ for $i=1,\dots,n-1$, starting at the current value of the fast parameters $x=x_0$ and ending at a final value $x'\equiv x_{n-1}$. In practice this is done by making one or more Metropolis steps for the distributions $P_i(x)$.
\item Accept the entire move $(x,y) \rightarrow (x',y')$ with probability
\be
\min\left[ 1, \exp\left( \frac{1}{n}\sum_{i=0}^{n-1} \left[\ln P(x_i,y) -\ln P(x_i,y')\right]\right)\right].
\ee
\end{enumerate}
See Ref.~\cite{Neal04} for further details and the proof that the method satisfies detailed balance and hence converges to the desired distribution. Note that at the last step the chain is not sampling from $P(x|y')$, but the interpolating distribution  $P_{n-1}(x)$ which is one step away towards $P(x|y)$. For tightly correlated variables as shown in Fig.~\ref{dragging} the overall acceptance probability will only be high if the number of steps is high enough that  $P_{n-1}(x)$ is close to $P(x|y')$, and hence a large number of steps are required.

If the variables are uncorrelated, then $n=1$ will work well, which is equivalent to a standard Metropolis step with no interpolating distributions. It may be advantageous to make one or more additional standard Metropolis steps from $P(x|y')$ after each dragging step, but any difference in performance will be small for large numbers of interpolating distributions. For uncorrelated fast and slow parameters a large $n$ allows the fast parameters to move to a new random position for each slow move, and when there are correlations between fast and slow parameters it allows the fast parameters to move to the region of high likelihood at the new slow position.

Given that the current value of $P(x,y)$ is known, the entire dragging step only requires one new slow evaluation at the new value  $y'$. Whether the method is helpful in practice depends on whether all the fast likelihood evaluations for $P(x_i,y)$ and $P(x_i,y')$ are fast enough that the additional computational cost is worth it for improved movement in the slow parameter subspace. Note that the method requires $2n$ fast likelihood evaluations per slow evaluation.

There are free parameters in the method determining the number of intermediate distributions to sample from $n$, and the number of intermediate Markov chain steps at each intermediate distribution. Typically the number of interpolating distributions required will depend on the number of fast dimensions, and I define $\ndrag$ so that $n= \ndrag\nfast$ (see e.g.~\cite{Neal94samplingfrom}, though in general the optimal number of interpolating steps will depend in a unobvious way on the structure of the distribution being sampled). Since the method benefits from the interpolating distributions being as close to each other as possible \COSMOMC\ does just one Metropolis step per interpolating distribution. There is also a choice over which parameters to drag; for the scheme to be efficient, there needs to be good hierarchy in the parameter speeds. \COSMOMC\ groups fairly slow parameters together, and treats these as `slow' variables, and groups all the faster parameters together and drags them. It may be possible to devise a more general hierarchical dragging scheme, but it would require a good hierarchy between all the blocks and is not well motivated by current applications at this point. If there are fast parameters which are known to be nearly independent to other parameters these could be separated out and sampled in their own standard Metropolis steps to avoid the computational overhead of unnecessarily dragging them.

\section{Performance comparison}
\label{sec:performance}

The performance of different sampling methods can vary significantly between problems, and there is no general `best' method.
I start by briefly considering a few simple limiting cases where the behaviour of the sampling methods is easily understood, and then go on to discuss indicative convergence statistics for various fast-slow sampling methods when applied to sampling for cosmological parameter analysis.

\begin{itemize}
\item
\emph{Independent fast and slow parameters:}
The speedup from separating the fast and slow proposals here is $\clo(1 + \nfast/\nslow)$. The advantages to oversampling the fast parameters ($\oversample>1$) in this trivial case are only the following: (1) greater sampling accuracy on the fast parameters; (2) slight reduction the variance of any calculations involving fast parameters; and (3) more rapid convergence if the fast parameter subspace is harder to sample than the slow subspace. The dragging scheme has no advantages in this case and an overhead that makes it less efficient.
\item
\emph{Gaussian posterior with correlated fast and slow parameters:} When the covariance is known accurately a priori, the Cholesky parameter rotation will completely decorrelate the parameters and for Gaussian parameters also make them independent. This case is then equivalent to the first case. However if the covariance is not known a priori, a dragging scheme may be much more efficient than a Metropolis scheme because it allows the slow parameter space to still be explored almost as though the fast parameters had been marginalized out. In practice, an adaptive scheme that learns the covariance will be even more efficient (such as that implemented in \COSMOMC; see Sec.~\ref{adaptive} below).
\item
\emph{Modest speed hierarchy:} The Cholesky decorrelation scheme will allow relatively modest speed hierarchies to be exploited. In this case  $\oversample\approx 1$ is likely to be optimal unless the faster space is significantly harder to sample. However for an  ${\cal O}(1)$ speed hierarchy the dragging scheme is likely to be significantly slower due to the likelihood overhead and the imposed asymmetry between the treatment of the different parameters. Also, in the limit that all parameter speeds are very nearly the same it may be better not to use a fast-slow method at all, so that random rotations of the proposal explore all possible directions.

Note that if there are two likelihood components with identical internal speed but some common nuisance parameters, there is still a speed hierarchy: when updating the shared parameters both likelihoods must be reevaluated, but updating the independent parameters only one likelihood changes, so the common shared parameters are twice as slow as the independent parameters.
\item
\emph{Very fast parameters:} In the case that the computational cost of fast steps is always negligible,  the dragging method will always be more efficient since it becomes equivalent to sampling in the fast-marginalized slow subspace (except in the special case where fast and slow parameters are independent).
\end{itemize}

Note that the correlation lengths of the output chains are quite different in the different sampling methods. The dragging scheme only generates new samples where changes in the slow parameters are proposed, but the Metropolis schemes will generate many more samples including steps that only involve fast parameters. For independent fast and slow parameters the dragging method output is roughly equivalent to taking the Metropolis output and thinning it by a factor $1+\oversample\nfast/\nslow$. \COSMOMC\ automatically thins Metropolis runs by a factor $\oversample$ so that large $\oversample$ can be used without making output files very large and avoiding possible additional disk access overheads.

\subsection{Quantifying convergence}

There are many ways to test and quantify chain convergence, all of which are necessary but not sufficient to guarantee correct answers. Here I quantify chain convergence using a single simple generalized Gelman-Rubin statistic~\cite{Gelman92,Brooks98}, $R-1$. This measures the variance in parameter means evaluated from different chains in units of the parameter variance, which should be a small number for well-converged chains so that the posterior means are accurately measured. In detail, using all of the post burn-in samples from $n$ chains, where each sample gives a vector of parameter values $\vx$, this is evaluated with the following steps:
\begin{itemize}
\item Calculate the mean $\bar{\vx}_i$ from each chain, and the total mean from all chains $\bar{\vx}$.
\item Calculate the covariance between $n$ chains of the chain means
$$\mC_{\bar{\vx}} \equiv \frac{1}{n-1}\sum_{i=1}^n (\bar{\vx}_i-\bar{\vx})(\bar{\vx}_i-\bar{\vx})^T.$$
This should be small compared to the parameter covariance if the chains are all well converged.
\item Calculate the mean of the covariances within each chain
$$\mC_{\vx} \equiv \frac{1}{\sum_i w_i}\sum_{i=1}^n w_i \langle({\vx}-\bar{\vx}_i)({\vx}-\bar{\vx}_i)^T\rangle$$
where angle brackets denote the weighted mean within the chain and $w_i$ is the sum of the weights in each chain.
\item Cholesky decompose the mean covariance $\mC_{\vx} = \mL \mL^T$ to orthonormalize the parameters by forming $L^{-1}\vx$.
\item Calculate the eigenvalues $D_i$ of the between-chain covariance of the orthonormalized parameter means using
$\mL^{-1} \mC_{\bar{\vx}} [\mL^{-1}]^T = \mU \mD \mU^T$.
\item Define $R-1\equiv \max(D_i)$ to measure the largest variance between chains of any of the orthonormalized parameter means.
\end{itemize}

The $R-1$ statistic can be evaluated for all parameters or just a subset of particular interest. Specifically I shall compare the values obtained from just the slow parameters with the value for all of the (fast+slow) parameters. Convergence in the slow subspace can be significantly better than in the full space if there are only weak fast-slow dependencies and the fast space is hard to explore (and vice versa). Requiring convergence on the full space is always more conservative, and typically chains are run to obtain $R-1\alt 0.1$. Smaller values may be required if dense sampling is required for estimating two and three-dimensional densities, robust evaluation of confidence intervals, or later importance sampling. Often aiming for $R-1\sim 0.02$ works well for many purposes. If confidence intervals are the primary concern the Gelman-Rubin statistic can also be generalized to test convergence of the between-chain variance of the confidence intervals (rather than the default measure of the covariance of the means).

\subsection{Cosmological parameters with \planck}

\begin{figure}[t]
\begin{center}
\includegraphics[trim=0.5cm 0 0 0,width=9cm]{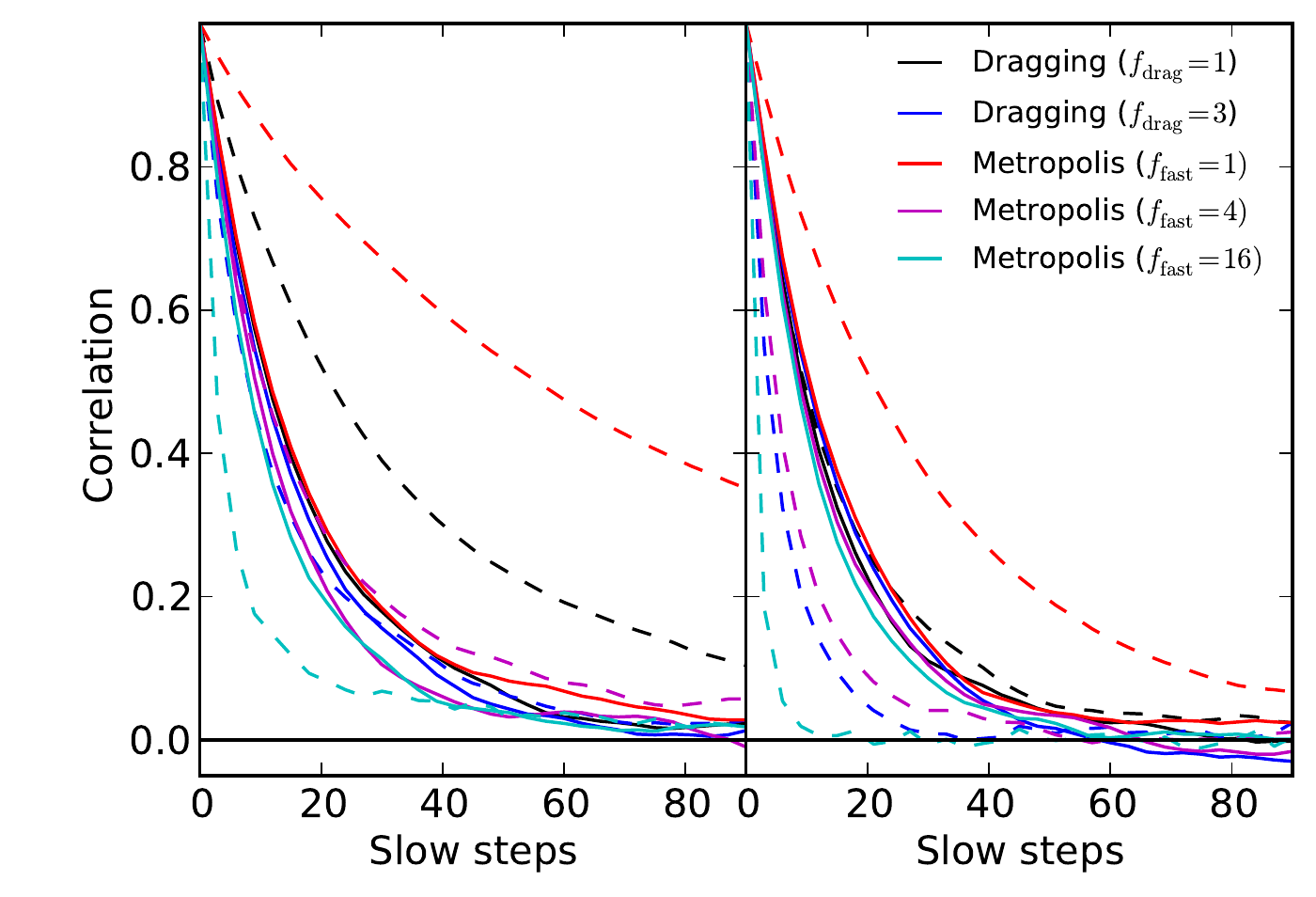}
\caption{Indicative correlation length in units of slow parameter steps for \planck\ parameter estimation runs in the baseline six-parameter model with known covariance. Left shows \planck+WP ($\nfast=13$), right shows \planck+WP+highL ($\nfast=31$). Solid lines show the correlation length for one of the slow parameters, dashed lines show one of the fast parameters. Using the dragging method with $\ndrag \ge 2$ or Metropolis with $\oversample\agt 4$ reduces the fast-parameter correlation length to be similar or smaller than the slow correlation length, improving convergence compared to the simplest fast-slow scheme (Metropolis with $\oversample=1$). Actual total efficiency depends on the relative speed of the fast and slow steps.
\label{corrLength}}
\end{center}
\end{figure}

\begin{table*}
\centering
\begin{tabular} {| l || c | c|}
\hline
  \multicolumn{3}{|c|}{\planck+WP ($\nslow=6,\nfast=13$)} \\
\hline
Method & All ($R-1$)  &  Slow  ($R-1$) \\ [0.5ex]
\hline
\hline
No fast-slow &0.27 & 0.10\\
\hline
Dragging $\ndrag=1$  & 0.060 & 0.016  \\
Dragging $\ndrag=2$  & 0.025 & 0.008\\
Dragging $\ndrag=3$  & 0.029 & 0.011\\
Dragging $\ndrag=5$  & 0.018 & 0.014\\
\hline
Metropolis $\oversample=1$ & 0.065 & 0.020  \\
Metropolis $\oversample=2$ & 0.035 & 0.008  \\
Metropolis $\oversample=4$ & 0.025 & 0.005  \\
Metropolis $\oversample=8$ & 0.018 & 0.011  \\
Metropolis $\oversample=16$ & 0.011 & 0.007  \\
\hline
\end{tabular}
\quad\quad
\begin{tabular} {| l || c | c|}
\hline
  \multicolumn{3}{|c|}{\planck+WP+highL ($\nslow=6,\nfast=31$)} \\
\hline
Method & All ($R-1$)  &  Slow  ($R-1$) \\ [0.5ex]
\hline
\hline
No fast-slow & 0.25 & 0.074\\
\hline
Dragging $\ndrag=1$  & 0.031 & 0.010  \\
Dragging $\ndrag=2$  & 0.020 & 0.009\\
Dragging $\ndrag=3$  & 0.023 & 0.008\\
Dragging $\ndrag=5$  & 0.027 & 0.027\\
\hline
Metropolis $\oversample=1$& 0.042  & 0.004  \\
Metropolis $\oversample=2$& 0.015 & 0.003  \\
Metropolis $\oversample=4$& 0.014 & 0.008  \\
Metropolis $\oversample=8$& 0.012 & 0.008  \\
Metropolis $\oversample=16$& 0.011 & 0.008  \\
\hline
\end{tabular}

\caption{
Typical convergence values achieved using various fast-slow sampling methods after fixed wall time for the case study of six baseline cosmological parameters from \planck\ data, including also WMAP polarization (WP). The left table is for a five-hour run with a relatively unconstrained and highly degenerate foreground (fast nuisance parameter) model with $\nslow=6,\nfast=13$. The right table shows results from an eight-hour run also using additional data  (highL) that constrain the foreground model better, making it less degenerate and more Gaussian, but adding 18 additional fast nuisance parameters. The columns give Gelman-Rubin $R-1$ values evaluated for the least-converged direction in either the full parameter space (All) or the slow parameter subspace (Slow - which are the parameters of most physical interest in this case). The precision quoted is higher than sampling fluctuations, so exact numbers should not be over-interpreted, but the great improvement over the non-fast-slow method is clear, and also the advantage of using $\oversample>1$ for Metropolis sampling. In all cases an accurate  precomputed fixed covariance matrix was used to orthogonalize the parameters. The fast and slow parameters are well decorrelated, so the overhead of the dragging method makes it less efficient than $\oversample\gg 1$ with Metropolis sampling in this case.
}
\label{planck_converge}
\end{table*}

The \planck\ satellite provides high-resolution maps of the microwave background and corresponding likelihood functions. In addition to the theoretical power spectrum (a function of spherical wavenumber $l$), the likelihood also depends on a number of calibration, beam mode, foreground amplitude and foreground correlation `nuisance' parameters as described in detail in Ref.~\cite{Ade:2013lta}. The latter parameters are important to model the small-scale data reliably where foreground contamination and beam errors are non-negligible. However in this regime the likelihood is also well described for most theoretical models by a fiducial Gaussian approximation, where one likelihood evaluation only requires a fast matrix vector multiplication using a precomputed inverse covariance matrix. Since the foreground models are all very simple power laws or template amplitudes, when any of the nuisance parameters are changed only the high-$l$ part of the likelihood needs to be recomputed, not the theoretical power spectra nor the low-$l$ likelihood. The nuisance parameters are therefore fast parameters compared to the slow parameters determining the cosmological model which require a new calculation of the theoretical power spectra using~\CAMB\ (and also the low-$l$ polarization likelihood). For this test case the speed hierarchy is $\clo(100)$, so that $\clo(100)$ nuisance parameter changes can be made for the same computational cost as one change in the slow cosmological parameters.

Here I consider just two simple cases to give indicative performance. I follow the parameter estimation assumptions described in Ref.~\cite{Ade:2013lta}, and use the public \planck\ likelihood code\footnote{\url{http://www.sciops.esa.int/wikiSI/planckpla/}} in combination with \COSMOMC\ (which uses \CAMB\ for the theoretical calculation~\cite{Lewis:1999bs}). The two cases have differing properties and numbers of fast parameters: ``\planck+WP" is where the likelihood only includes \planck\ data and polarization from WMAP (WP;~\cite{Bennett:2012fp}); ``\planck+WP+highL" is an extended data combination which includes data from other high-$l$ CMB observations (highL; \cite{Reichardt:2011yv,Das:2013zf}). The \planck\ likelihood depends on 13 nuisance parameters, the highL likelihood on 24, but 6 of these are determining physical foregrounds that are common to the \planck\ likelihood. There are therefore a total of 13 and 31 different fast parameters in the two cases, with one fast block in the \planck+WP case and two fast blocks (taken to be of 13 and 18 parameters each) in the \planck+highL case. The WMAP polarization likelihood has no nuisance parameters and does not need to be recomputed unless the slow cosmological parameters change.

The underlying cosmological model is usually described by six or more slow cosmological parameters\footnote{For speed I will use zero neutrino mass here, which slightly favours fewer fast steps compared to a realistic calculation.}.
For the theoretical calculation there are two blocks of slow parameters: general cosmological parameters that require full reevaluation of the linear transfer functions, and semi-slow parameters than determine the initial conditions as parameterized by the initial power spectrum. In itself there is a significant speed difference between these blocks, but both sets of parameters also require the low-$l$ likelihood to be reevaluated using the theoretical prediction, and this step can have a non-negligible computational cost that makes the total hierarchy in speed rather more modest. This speed hierarchy can still be exploited using two blocks of Cholesky decorrelated parameters, but it is not large enough for a dragging method to be efficient.

The convergence for given wall time depends on the method used and the relative speed of the fast and slow calculations. There is a trade-off between having more fast steps --- which improves movement in the slow parameter space for the dragging scheme, and in general speeds convergence of the fast subspace --- and slowing things down by having so many fast steps that their computational cost becomes significant. In all test cases I ran four chains (on one node, with four cores per chain, 16 cores in total) with a variety of sampling methods, and used a proposal distribution scale of 2 (see Appendix~\ref{proposal}).

\subsubsection{Baseline case with known covariance}

First I use an accurate precomputed covariance matrix for the parameter decorrelation, which will tend to favour standard Metropolis over dragging schemes, and sample the six parameters of the baseline \LCDM\ cosmology. The $R-1$ convergence statistics achieved after fixed wall time are summarized in Table~\ref{planck_converge} for various sampling configurations. Corresponding chain correlation lengths are shown in Fig.~\ref{corrLength} as a function of the number of slow steps along the chain.

All the fast-slow methods considered significantly outperform the simple non-fast-slow Metropolis method, with speedup $\clo(1+\nfast/\nslow)$. The fast parameter space is actually less Gaussian than the slow space (e.g. due to various hard parameter priors), and hence harder to explore. The fast-slow method speed-up can actually be better than $1+\nfast/\nslow$ since as shown in Fig.~\ref{corrLength} it is the fast parameters that determine the overall correlation length; the fast-slow sampling schemes (for $\oversample\gg1$) can reduce this to being less than the slow parameter correlation lengths and hence give additional gains. This is mitigated by the non-negligible cost of the fast steps, so that eventually performance becomes worse as vastly more fast calculations are required per slow step (the dragging method with $\ndrag=5$ is clearly worse in the 37-parameter case than $\ndrag=3$).

For the fast-slow Metropolis schemes the convergence in the full space is significantly improved by having $\oversample \gg 1$. Since the fast and slow parameters are decorrelated there is actually a trade-off between better convergence in the fast subspace and the slow subspace: for  $\oversample\agt 4$ convergence in the slow parameter space becomes worse due to the lower total number of slow steps that can be performed in the given computation time, but larger $\oversample$ continues to improve fast subspace convergence.
The dragging methods achieve good convergence in the full and slow subspaces for $\ndrag \agt 2$, though they are outperformed in this case by the simpler Metropolis scheme since there is only a modest gain in slow subspace movement once the parameters are fully decorrelated. Convergence in the fast subspace could also be improved by using a proposal distribution more tailored to the non-Gaussian shape of the fast-parameter subspace.

The best fast-slow parameter choices seems to perform well, with convergence achieved for fixed number of slow likelihood evaluations being similar to that expected for sampling with no nuisance parameters. The speed hit from the nuisance parameters is therefore limited to the numerical cost of calculating the fast steps, which for good parameter choices does not dominate the numerical cost of the slow step; the overall efficiency is then within $\clo(1)$ of the performance expected for no fast nuisance parameters.


\subsubsection{New parameters with unknown covariance}
\label{adaptive}
%

In reality the covariance is also often unknown a priori, and a common situation is testing a new model with new parameters (with unknown correlations) in addition to the six parameters of the baseline \LCDM\ model, or changing the number and modelling of the nuisance parameters.  In these cases it is usually highly beneficial to do an initial run to estimate the covariance, or use an adaptive scheme that gradually learns the covariance, with decorrelation of the parameters in subsequent steps. \COSMOMC\ uses an adaptive scheme that uses a growing fraction of the previous chain samples to estimate the covariance used for the proposal distribution.

The adaptive method is asymptotically valid as long as the number of samples used to estimate the correlation grows as a fixed fraction of the total samples so the covariance itself converges, and hence the subsequent updates are Markovian.
While the covariance estimate is changing significantly the process is strictly non-Markovian, and these early fraction of steps can be removed as extended burn in. \COSMOMC\'s algorithm is essentially a version of the adaptive Metropolis method~\cite{Haario01}, with the covariance being updated periodically from the average of multiple chains for speed of MPI implementation. Since the proposal is only updated periodically the method is also piecewise Markovian in the limit that the exploration time is short compared to the update time.
For discussion of ergodicity of adaptive Metropolis methods see Refs.~\cite{Haario01,2009arXiv0911.0522V} and references therein.

As a simple test case I'll consider a simple approximation to adding a new set of fast parameters with unknown covariance. This is a common situation: for example, before the \planck\ nuisance-parameter model was defined the expected cosmological parameter covariance was known from forecasts, but the covariance with the new foreground model was unknown. To avoid wasting computer time on simple tests, I approximate the full \planck\ likelihood as Gaussian, with covariance given by that from the actual full run, and then do test runs starting with a diagonal fast-parameter covariance set to some initial guess at appropriate widths.

First consider the case with non-adaptive sampling, so the proposal distribution is fixed from the beginning (and fast and slow parameters are not decorrelated because the full covariance is unknown). In this case the dragging method can perform significantly better, though here the fast and slow parameters are sufficiently weakly correlated that dragging performs similarly to Metropolis sampling with high $\oversample$ and there is no clear advantage. In extended models with stronger correlations between fast and slow parameters the dragging scheme is expected to perform better.

However the adaptive scheme that learns the covariance and in subsequent steps uses it to define the proposal distributions and fast-slow decorrelation is around twice as efficient for \planck+WP, and around three times as efficient for \planck+WP+highL, so there is a clear advantage to using an adaptive method. This gain would be even larger if new correlated slow parameters were also being added.
The adaptive proposal learning scheme is straightforwardly combined with any of the fast-slow sampling methods, and is recommended unless a good covariance is known a priori.

\subsection{Likelihood requirements}

As we have seen, significant performance gains are possible using fast-slow sampling methods. For this to work it is essential that all parts of the likelihood calculation that have different speed and parameter dependence are separated. In some cases this may require a significant amount of additional thought into how to structure the calculation. If the computational cost of changing the fast parameters is small but non-negligible, there will be additional gains to be had by optimizing the fast likelihood, even though in a non-fast-slow method it would contribute insignificantly to the total numerical cost.

\section{Conclusions}

Using efficient fast-slow sampling methods can significantly improve the performance of parameter inference when there are large numbers of fast parameters. I've demonstrated that a simple speed-ordered Cholesky orthogonalization can provide substantial performance gains for problems currently of interest in cosmology. This can be combined with an adaptive covariance learning scheme and/or the fast-parameter dragging method to improve robustness and performance when there are non-trivial correlations or dependencies between the fast and slow parameters. Although I have focussed on application in cosmology, the algorithms are general, and a wide class (though certainly not all) problems are likely to have a similar speed structure.

I have only considered here the case of parameter estimation, but similar efficiency gains should also be achievable for evidence (free energy) calculations. Algorithms such as thermodynamic integration generalize straightforwardly to exploit fast and slow parameters, other algorithms however may require more work to adapt and this should be the subject of further investigation. A variety of other general fast-slow sampling methods have also been proposed by Ref.~\cite{Neal11}, which may give performance gains in some problems.

  The focus of this paper in on direct fast-slow sampling schemes for generating samples from the full posterior. However, in particular cases, there may of course be better alternatives. For example it may be possible to accurately approximate the slow part of the likelihood. If the approximation can be made accurate enough (without an over-expensive precomputation step), then full chains can simply be run directly. For example in the cosmological context the \PICO\ approximation~\cite{Fendt:2007uu} can be used to calculate the CMB power spectra quite accurately, which can save a lot of time if the goal is to explore many different fast nuisance parameter models with the same underlying set of cosmological models. In other cases the approximation may not be accurate enough, but still a sufficiently good approximation that later importance sampling correction from the exact likelihood works well (see e.g.~\cite{Lewis:2002ah}). This can be a good solution if the full likelihood is too slow for a fast-slow direct sampling method to be useable, and allows rapid full exploration in the fast-approximated likelihood space; only the highly-thinned independent samples then need to be corrected by later importance sampling, a step that is trivially parallelizable. In cases where importance sampling cannot be used efficiently there are also sampling schemes that can exploit fast approximations, see for example Ref.~\cite{Wang13} and references therein.

The numerical Fortan 2003 code \COSMOMC\ discussed here is publicly available\footnote{\url{http://cosmologist.info/cosmomc/}}, which implements both standard Metropolis and dragging sampling methods, along with the \GetDist\ program for analysing samples and generating marginalized 1, 2 and 3-D posteriors, and python scripts for managing and running grids of runs, processing the results, and making latex tables and plots (as used in the main \planck\ parameter analysis of Ref.~\cite{Ade:2013lta}).

\section{Acknowledgments}
I thank Sarah Bridle for collaboration on developing the \COSMOMC\ code.
I thank fellow members of the \planck\ collaboration for many related useful discussions, and members of the community for contributions to debugging and development of \COSMOMC.
I acknowledge support from the Science and Technology Facilities Council [grant number ST/I000976/1].

\appendix

\section{Choice of proposal distribution}
\label{proposal}

Metropolis sampling methods work with any symmetric proposal distribution in
principle, as long as it allows eventual exploration of the full
parameter space. Gaussian proposal distributions are often used, but these are not necessarily optimal.

As discussed in the text, it usually pays to transform to orthogonalized
parameters for efficient parameter exploration. The fast-slow Cholesky parameter redefinition achieves this, but it remains to make a choice of proposal distribution in the orthonormalized parameter space.

\begin{figure}
\begin{center}
\includegraphics[width=\hsize]{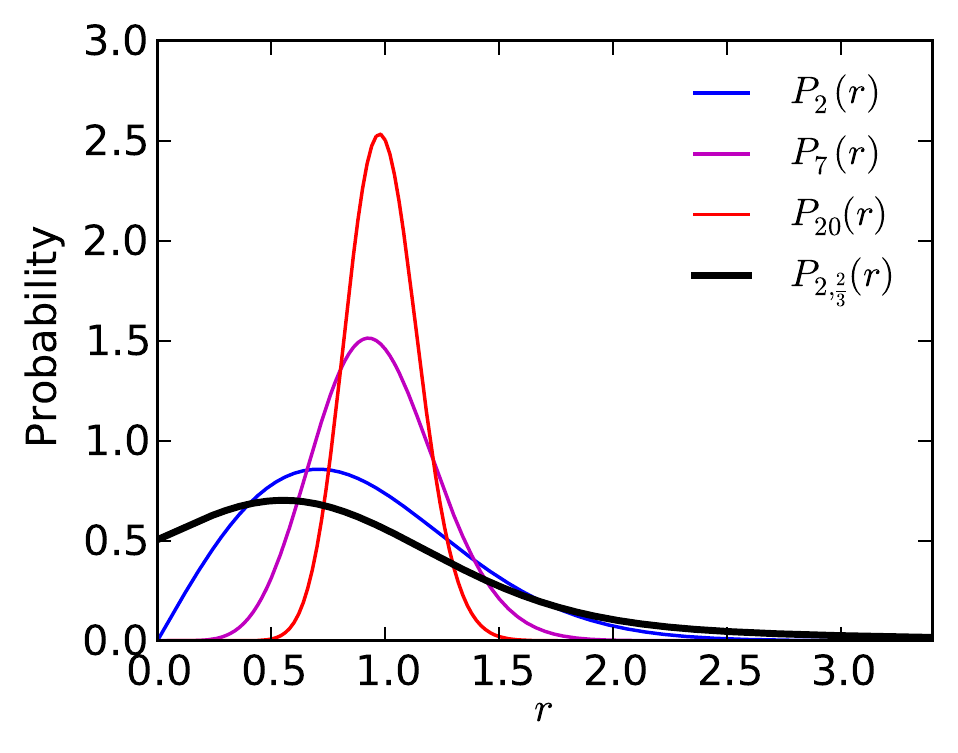}
\caption{Possible radial proposal distributions. An $n$-D Gaussian proposal distribution corresponds to choosing a random direction in parameter space and proposing a move by distance $r$ in that direction with probability $P_n(r)$; larger $n$ become more sharply peaked near one, and in particular very rarely propose much smaller or larger moves. The thick black line $P_{nf}(r)$ shows a mixture with fraction $f=\frac{2}{3}$ of $P_2(r)$ and fraction $1-f$ of an exponential distribution. This has much broader tails and does not go to zero at $r\sim 0$, and is the proposal distribution used by \COSMOMC. It is much more robust to covariance matrix misestimation than a high $n$-D Gaussian proposal distribution (though slightly less optimal in the ideal case).
 \label{propose_r}}
\end{center}
\end{figure}

First consider the case of sampling an orthonormalized parameter space using a Metropolis method with an $N$-dimensional Gaussian proposal distribution.
Sampling from an $N$-D Gaussian proposal distribution amounts to choosing a direction at random
(a direction on the surface of an $N-1$ sphere), then making a
proposal $P_N(r)$ for the distance $r$ along that direction with
\be
P_n(r) \propto r^{n-1} e^{-n r^2/2}.
\ee
The distance may be scaled by a factor $s$, and
Ref.~\cite{Gelman95,Dunkley:2004sv} show that $s\sim 2.4$ is optimal in terms
of minimizing the correlation length when the target distribution is Gaussian. I define $P_n^s(r) \propto
P_n(r/s)$ so that the optimal radial proposal distribution for the choice of an $N$-dimensional Gaussian in $P_N^{2.4}(r)$.

One simple improvement on this is to cycle directions rather than
choosing a random direction for each move. There are $N$ orthonormal
eigendirections that make a basis for the space, but proposals along these directions can be made in any order. One good option is to
 first choose a random basis rotation, then cycle through the
basis, making proposals along each basis vector direction in turn. When all
the basis vector directions have been proposed, generate a new random
basis and repeat. This cycling avoids random directions sometimes heading back
where they have just come from, and helps to remove random noise from
the number of proposals in different directions. This can improve exploration especially in relatively low dimensions.
%
%
%
%

The choice of an $N$-dimensional proposal distribution may also be suboptimal.
Consider separating the choice of direction to move in and distance to
move. For example we could try some radial proposal function $P_n(r)$
for $n\ne N$, which would correspond (for $n<N$) to making a Gaussian proposal in an $n$-D subspace.
A good and robust proposal distribution
performs well in the ideal case, but also doesn't perform too badly if
the proposal width is not quite correct (for example if it has been
estimated from a fairly short initial sampling period). In this respect $P_2^{2.4}(r)$ is
significantly superior to $P_N^{2.4}(r)$ when sampling number of dimensions $N>2$: the distribution is
significantly less peaked, and therefore much less sensitive to the width
being chosen incorrectly. See Fig.~\ref{propose_r}.

Consider for example the case of a factor four overestimation
of the proposal width: sampling a 7D Gaussian with $P_2^{2.4}(r)$ the
autocorrelation at 50 steps is $C_{50}\sim 0.45$, whereas for  $P_7^{2.4}(r)$
it is much worse at $C_{50} \sim 0.9$. For factor 4 underestimation $P_7^{2.4}(r)$
performs only slightly better. For factor 8 overestimation  $C_{50}\sim 0.8$ when using $P_2^{2.4}(r)$, which is slow but not completely
useless, however $P_7^{2.4}(r)$ performs catastrophically badly because it almost never proposes small moves. For the ideal Gaussian case when
the proposal width is matched, $P_7^{2.4}(r)$ gives $C_{10} \sim 0.3$ as
opposed to $P_2^{2.4}(r)$ which gives the slightly less optimal correlation length $C_{10} \sim 0.42$. This cost may however be worth paying for more robustness. Also note that decorrelation is not necessarily a good indicator of efficient exploration of the full parameter space; for example a wide fixed-width proposal can often flip between tails on opposite sides of the mean and give rapid decorrelation, but leave the central region and more extreme tails  poorly explored (and hence poor overall convergence).

Underestimation of the proposal width can also be problematic and lead to slow random walk exploration. Using a proposal distribution with thicker tails helps with this. But underestimation is generally less problematic than overestimation because adaptive proposal updates (see Sec.~\ref{adaptive}) will gradually increase the proposal width to something more appropriate (whereas a large overestimation can lead to no chain movement in some direction, and hence no useful estimated covariance eigenvalue).

Having a broader proposal distribution is also likely to be
advantageous when the target is non-Gaussian, for example
containing broad tails or nearly-isolated local maxima.
The sampling can be made even  more robust to wrong proposal width estimation by increasing
the probability for small and large distance proposals.
One simple way to do this
is to mix in some component of an exponential distribution:
\be
P_{nf}(r) = f P_n(r) + (1-f) e^{-r}.
\ee
Taking $f \sim 2/3$, $n =2$ seems to work well, with performance
not much effected by the choice; this is the default choice in the \COSMOMC\ code. By design the efficiency is not very
sensitive to the proposal width, with a scaling of 1.5--2.5 generally working well and giving acceptance probabilities in the range 0.2--0.5. The distribution
shape is rather similar to using Gaussian proposals in random 1 to 3D
subspaces, but with somewhat broader tails at large $r$. For specific idealized cases this proposal distribution may be slightly suboptimal, but it is usually much more important for typical usage to perform well in most cases than to perform optimally in very specific test cases.

\bibliography{../antony,../cosmomc}

\end{document}